\documentclass[%
unsortedaddress,
preprint,
amsmath,
amssymb,
showkeys,
onecolumn]{revtex4-1}

\usepackage{graphicx}
\usepackage{dcolumn}
\usepackage{bm}

\usepackage{tcolorbox}
\usepackage{float}

\usepackage{tikz}
\newcommand*\circled[1]{\tikz[baseline=(char.base)]{
		\node[shape=circle,draw,inner sep=0.5pt] (char) {#1};}}

\begin{document}


\title{A kneading map of chaotic switching oscillations in a Kerr cavity with two interacting light fields }


\author{Rodrigues D. Dikand{\'e} Bitha}
 \email{rodrigues.bitha@auckland.ac.nz}
\affiliation{Dodd-Walls Centre  for Photonic and Quantum Technologies, New Zealand}%
\affiliation{Department of Physics, University of Auckland, Private Bag 92019, Auckland, New Zealand}%
\affiliation{Department of Mathematics, University of Auckland, Private Bag 92019, Auckland, New Zealand}%
\author{Andrus Giraldo}%
\affiliation{School of Computational Sciences, Korea Institute for Advanced Study, Seoul 02455, Korea}%
\author{Neil G. R. Broderick}
\affiliation{Dodd-Walls Centre  for Photonic and Quantum Technologies, New Zealand}%
\affiliation{Department of Physics, University of Auckland, Private Bag 92019, Auckland, New Zealand}%
\author{Bernd Krauskopf}
\affiliation{Dodd-Walls Centre  for Photonic and Quantum Technologies, New Zealand }%
\affiliation{Department of Mathematics, University of Auckland, Private Bag 92019, Auckland, New Zealand}%

\date{\today}

\begin{abstract}
Optical systems that combine nonlinearity with coupling between various subsystems offer a flexible platform for observing a diverse range of nonlinear dynamics. Furthermore, engineering tolerances are such that the subsystems can be identical to within a fraction of the wavelength of light; hence, such coupled systems inherently have a natural symmetry that can lead to either delocalization or symmetry breaking. We consider here an optical Kerr cavity that supports two interacting electric fields, generated by two symmetric input beams. Mathematically, this system is modeled by a four-dimensional $\mathbb{Z}_2$-equivariant vector field with the strength and detuning of the input light as control parameters. Previous research has shown that complex switching dynamics are observed both experimentally and numerically across a wide range of parameter values. Here, we show that particular switching patterns are created at specific global bifurcations through either delocalization or symmetry breaking of a chaotic attractor. We find that the system exhibits infinitely many of these global bifurcations, which are organized by $\mathbb{Z}_2$-equivariant codimension-two Belyakov transitions. We investigate these switching dynamics by means of the continuation of global bifurcations in combination with the computation of kneading invariants and Lyapunov exponents. In this way, we provide a comprehensive picture of the interplay between different switching patterns of periodic orbits and chaotic attractors.
\end{abstract}

 \maketitle


\section{Introduction}

Optical microresonators are chip-scale cavities that enhance and manipulate light.  In recent years, there has been a tremendous amount of theoretical and experimental work on Kerr microresonators driven by an external source~\cite{AGRAWAL1,lugi}. The dynamics of light in these cavities can be modeled by the Lugiato-Lefever equation~\cite{lugi}, which is a driven and dissipative nonlinear Schr\"odinger equation featuring loss, detuning, and sometimes gain.  Under specific conditions, this equation exhibits a wide range of dynamic behaviors, including bistability, chaos, and cavity solitons~\cite{cavsol1,cavsol2,cavsol3}. These are not only scientifically interesting but also have significant practical applications. Bistability is used for optical switching and memory in both communications and data storage applications~\cite{AGRAWAL1}, while chaos enables secure communications~\cite{REINISCH19941} and enhanced sensing technologies~\cite{chaos1}.  Cavity solitons have been utilized for generating optical frequency combs, which have revolutionized precision metrology, spectroscopy, and telecommunications~\cite{kippen1,Chembo1,kippen2}. 

Nowadays, applications of frequency combs are continually expanding. Presently, experiments often involve combining two or more frequency combs for purposes such as active sensing~\cite{sensdual} and dual spectroscopy~\cite{specdual}. A major challenge in these applications is maintaining a fixed relative phase relationship between the combs, especially when they originate from different laser sources~\cite{Schliesser2012}. One effective solution involves using optical systems based on either nested coupled microresonators or a single microresonator driven by multiple laser sources. Recent studies have shown that such setups can not only generate optical frequency combs and nested Kerr solitons with high efficiency but also achieve phase-locking across multiple resonators~\cite{chemboz}. A thorough understanding of these complex systems begins with the fundamental study of two-mode cavities, and they have recently captured the attention of numerous theoretical and experimental research works~\cite{kaplan,kaplan2,delbino2017,lewis2020,bitha}.

We consider here a ring resonator driven with two counter-propagating continuous waves with equal intensities, which is well described by the set of normalised coupled-mode equations:
\begin{equation}
	\begin{aligned}
		\frac{d E_1}{d t} &= \sqrt{F} -[1 + i ( \vert E_1 \vert^2 + B \vert E_2\vert^2 - \Delta)]E_1,
		\\
		\frac{d E_2}{d t} &= \sqrt{F} -[1 + i ( \vert E_2 \vert^2 + B \vert E_1\vert^2 - \Delta)]E_2.
		\label{eq1}
	\end{aligned}
\end{equation}
Here, $E_{1,2}$ represent the slowly varying complex amplitudes of the two counter-propagating modes in the ring resonator, and $F$ is the power of the input fields. The `slow' time, $t$, is measured in units of decay time, and $\Delta$ is the detuning from the nearest cavity resonance. The electric fields have been normalised such that the Kerr nonlinearity has a value of $1$, while the cross phase modulation has value $B$. Throughout this paper, we set $B = 2.0$; in this way, the dynamics of system~\eqref{eq1}  are qualitatively the same as the one observed in experimental realisation~\cite{delbino2017,selfswi}. System~\eqref{eq1} has an important symmetry property, namely $\mathbb{ Z }_2$-symmetry~\cite{equivariant,kuznesovbook}; more precisely, the system remains unchanged under the exchange mirror symmetry $\eta$ given by the transformation $\eta: (E_1,E_2) \rightarrow (E_2,E_1).$ Hence, the solutions of system~\eqref{eq1} -- steady states, periodic orbits, and chaotic attractors -- are either symmetric themselves (invariant under the transformation $\eta$) or asymmetric. Importantly, all asymmetric solutions come in pairs that are mapped to one another under $\eta$~\cite{bitha,equivariant,kuznesovbook,andrus2021}.

Experimentally, a platform modeled by system~\eqref{eq1} has been achieved by splitting the light of a narrow linewidth laser into two equal parts and then coupling each into a micron-size optical resonator in counter-propagating directions~\cite{delbino2017,selfswi}. System~\eqref{eq1} also models the dynamics of two orthogonally polarised light fields in a long fiber ring resonator~\cite{bruno2020} or in a Fabry-P\'erot resonator~\cite{Moroney2022}. Hence, the results presented here apply to all of these experimental systems.

This work follows from~\cite{bitha}, and we start by briefly summarising relevant previous results. Our main focus are the Shilnikov homoclinic bifurcations that are organized by a $\mathbb{Z}_2$-equivariant Belyakov transition.  We analyze and characterize the dynamics near this special point in parameter space, because it is crucial for understanding and predicting the emergence of chaotic attractors (and periodic orbits) of system~\eqref{eq1} with different oscillating behaviors in terms of which mode is dominant; we refer to these as switching patterns. To accomplish this, we analyze symbol sequences computed from the two output intensities; they encode the switching type and are known as kneading sequences. Computing and representing (in different colors) kneading sequences allows us to identify regions with the same switching pattern, and to show how their structure is related to the global bifurcations~\cite{andrus2021,barrio,barrio3}. We also compute the Lyapunov spectra of the two output intensities to determine the long-term stability of the different types of switching oscillations identified.

\section{Emergence of periodic orbits with different switching properties \label{ssnspo}}

Previous works have studied in detail the bifurcations of steady states~\cite{lewis2020,woodley2018} and \emph{non-switching} periodic orbits~\cite{bitha} as a function of the detuning $\Delta$. Building on this, we first investigate the bifurcations of system~\eqref{eq1} that give rise to periodic orbits with \emph{self-switching} oscillations. To achieve this, we employ the software AUTO07P~\cite{auto,kraus2007}, which allows us to follow steady states and periodic orbits from small to large detuning values, while keeping the input power $F$ constant.

\begin{figure}[]
	\centering
	\includegraphics[scale=0.9]{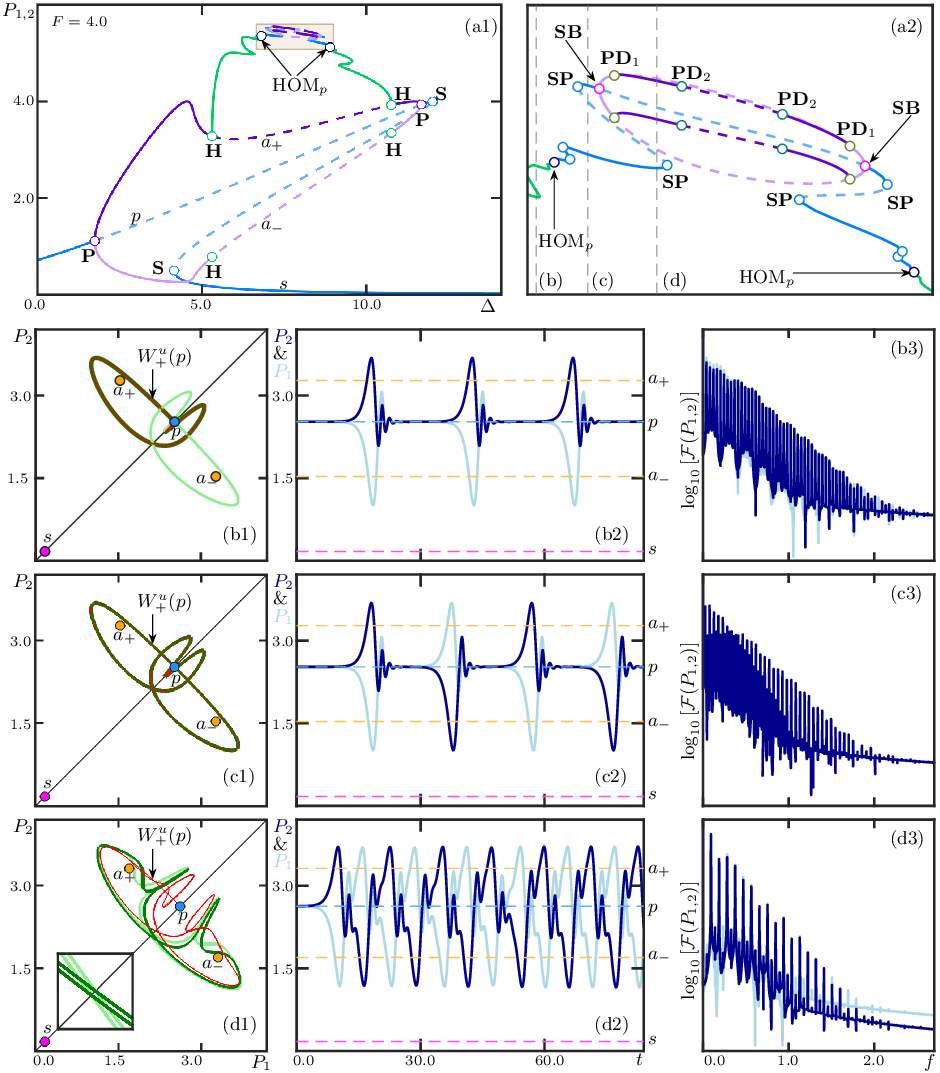}
	\caption{\label{fig:1dbif} One-parameter bifurcation diagram of system~\eqref{eq1} in the detuning $\Delta$, for $F=4.0$. Panel~(a1) is an overall view and panel~(a2) shows an enlargement in the colored frame. Stable and unstable symmetric solutions are represented by solid and dashed blue curves, respectively, and asymmetric states are shown in purple. Branches of periodic orbits (green) are represented here by the sum of the squared maxima of their real and imaginary parts, i.e., $P_{1,2}= \max(X_{1,2})^2 + \max(Y_{1,2})^2$. (c1)--(e1) Phase portraits with stable periodic orbits and the positive branch $W_+^u(p)$, (c2)--(e2) associated temporal traces, and (c3)--(e3) power spectra of the output intensities $P_1$ and $P_2$, for $(F,\Delta)=(4.0, 6.8053)$, $(F,\Delta)=(4.0, 6.8054)$, and $(F,\Delta)=(4.0, 7.16)$, respectively.}
\end{figure}

Figure~\ref{fig:1dbif} presents the one-dimensional bifurcation diagram of system~\eqref{eq1} as a function of $\Delta$ [row~(a)], along with selected phase portraits and temporal traces [rows~(b)--(d)], all for a fixed input power of $F=4.0$. At small detuning $\Delta$, the system exhibits a stable symmetric state, which means that the two output intensities $P_1 = |E_1|^2$ and $P_2 = |E_2|^2$ are equal. As $\Delta$ increases, the branch of symmetric steady states becomes unstable at a supercritical pitchfork bifurcation $\textbf{P}$; we denote this unstable (saddle) symmetric steady state $p$. Two stable branches of asymmetric steady states with $P_1 \neq P_2$, each being the counterpart of the other, emerge from  $\textbf{P}$. Thus, the point $\textbf{P}$ serves as a symmetry-breaking bifurcation~\cite{kaplan,equivariant,woodley2018}, where the system transitions from a symmetric to an asymmetric regime. By carefully selecting initial conditions, we can numerically determine whether $P_1$ or $P_2$ is dominant after the pitchfork bifurcation $\textbf{P}$. In contrast, this dominance occurs randomly in an experimental setting, due to the fact that it is very challenging to initial conditions. Thus, making this device a suitable candidate for random binary number generation~\cite{delbino2017, lquinn}.

Each asymmetric steady state undergoes two supercritical Hopf bifurcations $\mathbf{H}$, resulting in a detuning range where no stable asymmetric steady states exist. Rather, we find here a pair of asymmetric saddle steady states which we label $a_+$ and $a_-$. Following the second Hopf bifurcation $\textbf{H}$, the asymmetric steady states $a_{\pm}$   become stable again, until they both meet at a second pitchfork bifurcation $\mathbf{P}$, where they disappear and the symmetric steady state recovers its stability. Hence, this second bifurcation acts as a symmetry-restoring bifurcation~\cite{equivariant,fiedler2006global}. Thereafter, the symmetric steady state undergoes two saddle-node bifurcations $\mathbf{S}$, leading to another stable symmetric steady state with a low intensity, which we label $s$.

In system~\eqref{eq1}, Hopf bifurcations lead to the emergence of periodic orbits, which may then undergo further bifurcations, resulting in periodic orbits (and chaotic attractors; see already Sec.~\ref{phdiaggb}) with varying switching properties. More precisely, each pair of Hopf bifurcations $\textbf{H}$ generates a pair of asymmetric periodic orbits. Figure~\ref{fig:1dbif}(a) shows the corresponding branches originating from the two points $\textbf{H}$ on the branch of the dominant asymmetric steady state; these branches represent the maxima of $P_1$ and $P_2$ along the dominant periodic solution. Although not depicted here, these periodic orbits undergo sequences of period-doubling bifurcations, resulting in additional branches of periodic orbits with multiple loops~\cite{bitha, woodley2018}. The branches of asymmetric periodic orbits in Fig.~\ref{fig:1dbif}(a) that emerge from the points $\textbf{H}$ terminate at two distinct pairs of Shilnikov homoclinic bifurcations $\text{HOM}_p$, where system~\eqref{eq1} exhibits a pair of homoclinic orbits to the saddle-type steady state $p$; see already Fig.~\ref{fig:shilorb}(a). To compute these homoclinic orbits, we employ a two-point boundary value problem in combination with Lin's method~\cite{lin1}.

Numerical continuations indicate that the two bifurcations $\text{HOM}_p$ in Fig.~\ref{fig:1dbif}(a1) define a $\Delta$-range [in the light brown shaded region] within which periodic orbits undergo complex bifurcations. Previous research analyzed the dynamics in this $\Delta$-range by computing the Poincar{\'e} map of periodic orbits, revealing the presence of period-doubling  cascades~\cite{lewis2020,woodley2018}. Here, we wish to investigate whether these periodic orbits also exhibit other types of bifurcations; specifically the bifurcations that leads to periodic orbits with different switching properties~\cite{bitha, selfswi}. To this end, we compute the branches of periodic orbits emerging from both bifurcation points $\text{HOM}_p$. Figure~\ref{fig:1dbif}(a2) shows the enlarged one-parameter bifurcation diagram of periodic orbits within this region. For increasing $\Delta$ in Fig.~\ref{fig:1dbif}(a2), a new branch of periodic orbits emerges from the first Shilnikov homoclinic bifurcation point $\text{HOM}_p$. We find that periodic orbits along this new branch are symmetric.  Therefore, the Shilnikov homoclinic bifurcation $\text{HOM}_p$ serves as both a symmetry-increasing and symmetry-restoring bifurcation, where the solutions of system~\eqref{eq1} become symmetric again. The branch of symmetric periodic orbits undergoes two pairs of saddle-node bifurcations \textbf{SP}, creating $\Delta$-ranges  where two distinct periodic orbits coexist for the same parameter values. Subsequently, this branch encounters a pitchfork bifurcation \textbf{SB}, further confirming that the new periodic orbit is symmetric, as only symmetric solutions can bifurcate in this manner~\cite{equivariant,kuznesovbook}. Past the point $\mathbf{SB}$, the symmetric periodic orbit becomes unstable, and two branches representing the maxima of stable and symmetrically related periodic orbits emerge. As $\Delta$ is increased further, the branches of this new pair of asymmetric periodic orbits undergo period-doubling cascades, leading to and back from more complicated dynamics. Finally, the pair of asymmetric periodic orbits disappears at a second pitchfork bifurcation $\mathbf{SB}$, where the symmetric periodic orbit briefly regains stability before it ends at the second Shilnikov homoclinic bifurcation point $\text{HOM}_p$.

The bifurcations of the symmetric steady state $p$ are crucial in altering the dynamics of system~\eqref{eq1}. Therefore, analyzing its eigenvalues is essential. The saddle steady state $p$ has one positive eigenvalue $\lambda_u$, one negative real eigenvalue $\lambda_{ss}$, and a pair of complex conjugate eigenvalues $(\lambda_s, \lambda_s^*)$ with negative real parts. This eigenvalue configuration implies that the steady state $p$ has a one-dimensional unstable manifold $W^u(p)$ with two symmetry-related branches, $W_-^u(p)$ and $W_+^u(p)$, consisting of points that converge to $p$ in backward time. For $F=4.0$, the saddle value 
$$\sigma_p = \text{Re}(\lambda_s) + \lambda_u$$
of $p$ is always negative. Therefore, the point $\text{HOM}_p$ [in Fig.~\ref{fig:1dbif}(a)] represents a simple Shilnikov homoclinic bifurcation, from which a single periodic orbit bifurcates~\cite{shil2}.

Rows~(b)--(d) of Fig.~\ref{fig:1dbif} display periodic solutions within the $\Delta$-range shown in Fig.~\ref{fig:1dbif}(a2). More precisely, we showcase their phase portraits in the $(P_1,P_2)$-plane with the steady states $p$, $s$, and $a_{\pm}$, and the positive branch $W^u_+(p)$ of the unstable manifold of $p$. Additionally, we show their temporal traces of $P_1$ and $P_2$ and corresponding power spectra on a logarithmic scale. For $\Delta = 6.8053$, before the first point $\text{HOM}_p$, the positive branch $W^u_+(p)$ of the unstable manifold of $p$ accumulates on the top-left periodic orbit shown in Fig.~\ref{fig:1dbif}(b1), whereas $W^u_-(p)$ (not shown) accumulates on the bottom-right periodic orbit. These two periodic orbits form a pair of asymmetric solutions, related here by reflection across the diagonal $P_1 = P_2$ in the $(P_1, P_2)$-plane. The temporal trace of the top-left periodic orbit [Fig.~\ref{fig:1dbif}(b2)] reveals that $P_2$ is dominant most of the time. Therefore, we categorize the oscillations of this pair of periodic orbits as \emph{non-switching}. The power spectra [Fig.~\ref{fig:1dbif}(b3)] for this pair of periodic orbits show a Dirac comb (or frequency comb) with harmonics that are equally spaced but decreasing in amplitude; note the slight difference between the two spectra due to the fact that they are from a pair of asymmetric periodic orbits.

For $\Delta = 6.8054$, past the point $\text{HOM}_p$, the branch $W^u_+(p)$ now accumulates on a symmetric periodic orbit [Fig.~\ref{fig:1dbif}(c1)], which resembles the union of the two orbits from panel~(b1). Its temporal trace [Fig.~\ref{fig:1dbif}(c2)] indicates that both field intensities $P_1$ and $P_2$ oscillate around the asymmetric steady states $a_+$ and $a_-$. We can also observe regular intervals where each field intensity alternately dominates, resulting in \emph{self-switching} oscillations. The power spectra [Fig.~\ref{fig:1dbif}(c3)] of $P_1$ and $P_2$ along this symmetric periodic orbit are now the same and display odd harmonic teeth with smaller amplitudes, which are characteristic of self-switching oscillations. The temporal traces shown in panels~(b2)--(c2) indicate that, after each loop of oscillation, the branch $W^u_+(p)$ spends more time near the symmetric steady state $p$, resulting in periodic signals with a notably larger period  of oscillations. This phenomenon occurs because the parameter values of the corresponding periodic orbits are very close to the Shilnikov bifurcation point $\text{HOM}_p$, as highlighted in panel~(a2). For $\Delta = 7.16$, past the pitchfork bifurcation point $\textbf{SB}$, we observe the re-emergence of two symmetry-related periodic orbits, see Fig.~\ref{fig:1dbif}(d1). However, the temporal traces [Fig.~\ref{fig:1dbif}(d2)] associated with the top-left periodic orbit now exhibit two periodic signals with regular switching patterns and a slightly dominant $P_2$ intensity. These switching patterns are inherited from the dynamics of the previous symmetric periodic orbit, as illustrated in row~(c). In contrast to panels~(b2)--(c2), the temporal trace shown in panel~(d2) depicts two signals with a shorter period. This is because the parameter point for this temporal trace is sufficiently far from the point $\text{HOM}_p$. The power spectra of $P_1$ and $P_2$ [Fig.~\ref{fig:1dbif}(d3)] are now clearly different; moreover, they display fewer harmonic frequencies but a much larger repetition rate.

\section{Phase diagram of global bifurcations \label{phdiaggb}}

\begin{figure}[]
	\centering
	\includegraphics[scale=0.95]{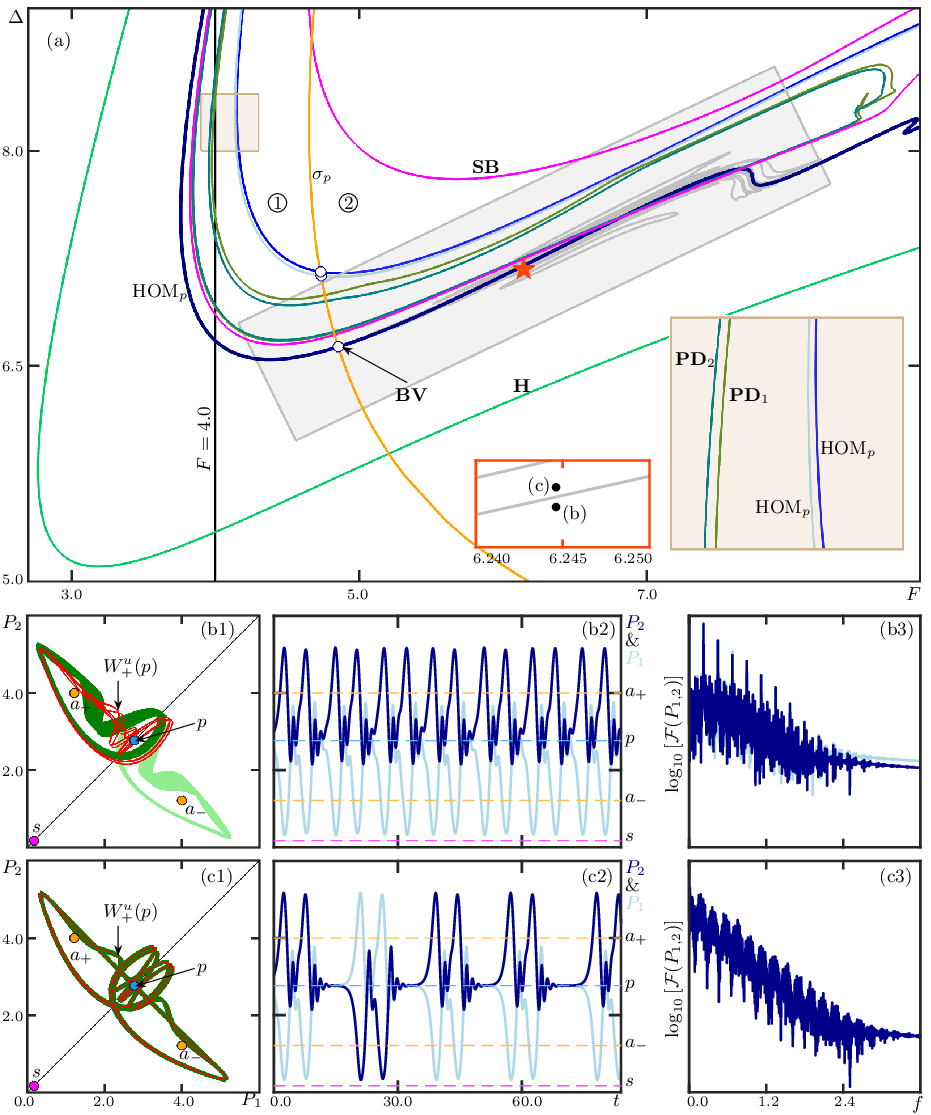}
	\caption{\label{fig:2dbif} (a) Phase diagram in the $(F, \Delta)$-plane showing a selection of bifurcation curves of system~\eqref{eq1}. Shown are the curve of Hopf $\mathbf{H}$ (green), Shilnikov  $\text{HOM}_p$ (dark blue, light blue and blue), pitchfork $\mathbf{SB}$ (fuchsia), and period-doubling $\mathbf{PD}_{1,2}$ (teal and dark green) bifurcations. The large and small insets show enlargements; in the colored frame in light brown and around the orange star, respectively. (b1)--(c1) Phase portraits with chaotic attractors and the positive branch $W_+^u(p)$, (b2)--(c2) associated temporal traces, and (b3)--(c3) power spectra of $P_1$ and $P_2$, for $(F, \Delta)=(6.2446,7.1740)$ and $(F, \Delta)= (6.2446,7.1753)$, as indicated in the small inset of panel~(a).}
\end{figure}

To map out the regions of existence for the various types of periodic orbits, we now employ numerical continuation methods to identify and follow their loci of bifurcations as the parameters $F$ and $\Delta$ vary. Figure~\ref{fig:2dbif} shows the resulting phase diagram of system~\eqref{eq1} in the $(F,\Delta)$-plane, together with phase portraits, temporal traces and corresponding power spectra at two selected parameter points. Our choice of the $(F,\Delta)$-plane is motivated by the fact that these two parameters are experimentally adjustable, while $B$ is usually fixed for a specific cavity.  Nonetheless, all the bifurcations discussed in this work can also be continued with respect to $B$. The phase diagram of system~\eqref{eq1} in the $(F,\Delta)$-plane, presented in Fig.~\ref{fig:2dbif}(a), highlights the loci of all the bifurcations observed in Fig.~\ref{fig:1dbif}(a), along with additional homoclinic bifurcations. The curve $\mathbf{H}$ in Fig.~\ref{fig:2dbif}(a) represents the locus of Hopf bifurcations, marking the boundary of the region where system~\eqref{eq1} exhibits periodic orbits (and chaotic attractors; see already rows~\ref{fig:2dbif}(b)--(c)) with varying switching properties. Within the area enclosed by the curve $\mathbf{H}$, we identify the loci of all the bifurcations of periodic orbits in Fig.~\ref{fig:1dbif}(a2), including the curves of the bifurcations $\text{HOM}_p$, $\mathbf{SB}$, $\mathbf{PD}_{1}$, and $\mathbf{PD}_{2}$, shown in dark blue, fuchsia, light green and teal, respectively. All these curves are located close to each other, and the large inset of Fig.~\ref{fig:2dbif}(a) provides an enlargement of the region shaded in light brown.

Numerical continuation reveals that the curve $\text{HOM}_p$ features a codimension-two Belyakov transition point $\mathbf{BV}$~\cite{bel,beltran2}, where the stability of the associated pair of homoclinic orbits changes. This change occurs because the steady state $p$ becomes a neutral saddle at the point $\mathbf{BV}$; specifically, the eigenvalues of $p$ satisfy the relation
$$\text{Re}(\lambda_s)+\lambda_u = 0.$$
As a consequence, the curve $\text{HOM}_p$ [Fig.~\ref{fig:2dbif}(a1)] is divided into two distinct sections, representing the locations of simple and chaotic Shilnikov bifurcations~\cite{shil2,shila,shill}. The first section of $\text{HOM}_p$, extending from the top part of the $(F,\Delta)$-plane to the point $\mathbf{BV}$, is the locus of simple Shilnikov bifurcations; here, the saddle value $\sigma_p$ is negative and a single periodic orbits bifurcates. The second section, from the point $\mathbf{BV}$ to the right, is the locus of chaotic Shilnikov bifurcations; here, the saddle value $\sigma_p$ is positive, meaning that the saddle steady state $p$ is more repelling than attracting.

When a curve of Shilnikov homoclinic bifurcation has a transition from simple to chaotic, a celebrated result by Belyakov~\cite{bel,beltran2} states that infinitely many curves of homoclinic bifurcations accumulate on one side of this primary curve. Moreover, when the steady state that exhibits this transition lies within the symmetry subspace of a $\mathbb{Z}_2$-equivariant system, additional curves of homoclinic bifurcations with different symmetry properties also exist on the other side of the primary homoclinic curve~\cite{bitha}. The gray curves in Fig.~\ref{fig:2dbif}(a) represent the loci of some homoclinic bifurcations resulting from the unfolding of the $\mathbb{Z}_2$-equivariant Belyakov transition point \textbf{BV}, as computed in~\cite{bitha}. 

We find that the steady state $p$ has zero saddle quantity along the curve $\sigma_p$ in Fig.~\ref{fig:2dbif}(a). This curve divides the $(F,\Delta)$-plane into two regions: in region~\circled{1}, the saddle quantity of the steady state $p$ is negative while, in region~\circled{2}, it is positive. Figure~\ref{fig:2dbif}(a) features two additional curves of homoclinic bifurcations (light blue and blue), whose associated orbits have a single loop of oscillation, similar to the primary homoclinic bifurcation $\text{HOM}_p$. The large inset of Fig.~\ref{fig:2dbif}(a) shows part of these two curves. Interestingly, the curve of neutral saddle $\sigma_p$ intersect the additional curves of homoclinic bifurcationsat the marked points, implying that each of them exhibits a Belyakov transition point as well. As we show later in Sec.~\ref{kneadbeli5}, these additional Belyakov transition points leads to infinitely more curves of homoclinic bifurcations.

We know that system~\eqref{eq1} exhibits simple dynamics in region~\circled{1}, as observed for selected parameter values shown in Fig.~\ref{fig:1dbif}(b)--(c). Now, we wish to confirm the existence of chaotic dynamics in region~\circled{2}, as predicted by Belyakov~\cite{bel,beltran2}. To this end, we present in rows~(b)--(c) of  Fig.~\ref{fig:2dbif}, the phase portraits, temporal traces, and power spectra of $P_1$ and $P_2$ for two values of $\Delta$, indicated in the small inset of Fig.~\ref{fig:2dbif}(a) and located on either side of a chaotic Shilnikov homoclinic bifurcation. For $\Delta=7.1740$ as in panel~(b1), the branch  $W^u_+(p)$ accumulates on a non-switching chaotic attractor. Notably, the presence of chaotic dynamics aligns with expectations near a chaotic Shilnikov homoclinic bifurcation. The temporal traces of $P_1$ and $P_2$ in panel~(b2) show two non-switching signals with a hint of period-two dynamics. Their power spectra are two different but similar combs with a few sharp harmonics and a complicated background of  harmonics with smaller amplitudes. We also note that each has a broadband frequency component, as expected for a chaotic attractor. Past the chaotic Shilnikov bifurcation, a symmetry-increasing bifurcation has occurred and, for $\Delta= 7.1753$ as in panel~(c1), we observe the formation of a symmetric chaotic attractor. Notice that the positive branch $W^u_+(p)$ of the unstable manifold of $p$ does several loops around the two asymmetric steady states $a_+$ and $a_-$. The temporal traces of $P_1$ and $P_2$ [panel~(c2)] not only reveal self-switching behavior and chaotic dynamics, but also display a hint of period-two dynamics, which is due to the nature of the nearby Shilnikov bifurcation. The power spectra of $P_1$ and $P_2$ of this symmetric chaotic attractor [panel~(c3)] now agree and are combs with a complex harmonic structure. The changes observed in Figs.~\ref{fig:1dbif}(b)--(c) and~\ref{fig:2dbif}(b)--(c) indicate that Shilnikov bifurcations of system~\eqref{eq1} are the mechanisms through which different types of symmetric periodic orbits and chaotic attractors appear or disappear --- with associated changes in the observed switching patterns of the output intensities.

\section{Mapping switching behaviors via kneading sequences \label{mapsecbel}}

We now dive deeper into how different switching patterns arise and where these are located in the $(F,\Delta)$-plane. To this end, we investigate how trajectories visit the two symmetry-related regions of phase space with $P_1 < P_2$ and $P_2 < P_1$, respectively. This is achieved with a symbolic dynamics approach, based on the so-called \emph{kneading sequence}, which encodes the switching pattern of trajectories with a sequence of symbols. This method has proven to be an effective tool to identify and delimit regions with different switching behaviors in parameter space~\cite{andrus2021,barrio,barrio3}.

The kneading sequence consists of symbols that describes how a trajectory visits different regions in the phase space. To generate the sequence for the specific case of system~\eqref{eq1}, we compute the positive branch $W_+^u(p)$ of the one-dimensional unstable manifold $W^u(p)$ of $p$ and consider how it oscillates around the asymmetric steady states $a_+$ and $a_-$, and also possibly the symmetric steady state $s$. To this end, we consider the associated time series of the difference $P_{\rm diff} = P_2 - P_1$. Each loop of $W_+^u(p)$ around $a_+$ corresponds to a large positive maximum of $P_{\rm diff}$, and we record the symbol $+$ for each such maximum; likewise, each loop of $W_+^u(p)$ around $a_-$ is characterized by a large negative maximum of $P_{\rm diff}$, and we record the symbol $-$. Here, large positive and large negative means that the respective maximum and minimum is outside the excluded range $[x^-,x^+]$, where $x^\pm$ are the values of $P_{\rm diff}$ at the steady states $a^\pm$; any smaller (in modulus) maxima and minima in between symbols $+$ and $-$ are ignored and do not generate a symbol $+$ or $-$. However, when there are no nearby attracting periodic orbits or chaotic attractors, $W_+^u(p)$ converges to the stable symmetric steady state $s$ and $P_{\rm diff}$ decays to $0$ in an oscillatory fashion along $W_+^u(p)$. In this case, we record the symbol $0$ for any positive maximum (or, alternatively, any negative minimum) to indicate that $W_+^u(p)$ makes a full loop around $s$. Note that this convention necessarily creates infinitely repeating symbols $0$, since the symbol $0$ can only be followed by another $0$.

For every typical (generic) point in parameter space, this procedure defines the infinite \emph{kneading sequence}
$$S_+=(s_1, s_2, s_3,s_4,\dots)_+ \quad \text{with} \quad s_i \in \{-,0,+\},$$ 
which represents the switching pattern of the output intensities. Because of the reflectional symmetry $\eta$ of system~\eqref{eq1}, the negative branch $W_-^u(p)$ generates the symmetric counterpart $S_-= \eta \left[ S+\right] = (\eta s_1, \eta s_2, \eta s_3, \eta s_4,\dots)$
with
$$
\eta s_i =
\begin{cases}
	+ & \text{if $s_i = -$},\\
	0 & \text{if $s_i = 0$},\\
	- & \text{if $s_i = +$}.\\
\end{cases}       
$$
Note that, by construction, the first symbol of $S_+$ is $+$, while that of $S_-$ is $-$. In what follows, we compute throughout the kneading sequence generated by $W^u_+(p)$ of $p$, which we will refer to simply as $S$ from now on. Moreover, we will denote infinitely repeating substrings of symbols with an overline; for example, $(++\overline{0})=(++0000\dots)$ and $(\overline{+-})=(+-+-+-\dots)$.

\begin{figure}	
	\centering
	\includegraphics[]{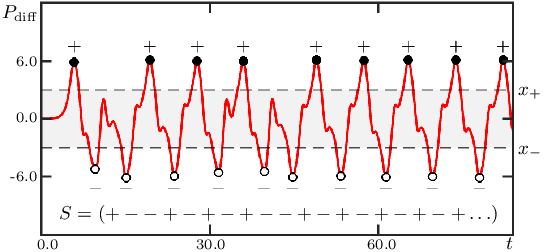}
	\caption{\label{fig:knead} Kneading sequence generated by the temporal trace of $P_{\rm diff}=P_2 - P_1$ along the positive branch $W^u_+(p)$ for  $(F,\Delta)=(7.78103,8.08573)$; the excluded range $[x^-,x^+]$ is shaded.}
\end{figure}

Figure~\ref{fig:knead} illustrates our definition of the kneading sequence for system~\eqref{eq1} by showing the temporal trace of $P_{\rm diff}$ along $W_+^u(p)$ for $(F,\Delta)=(7.78103,8.08573)$, which generate the kneading sequence $S=(+--+-+-+--+\dots)$. Notice in Fig.~\ref{fig:knead} how positive maxima and negative minima outside $[x^-,x^+]$ are recorded by the respective symbol, while those in this excluded range are not. As we observed in Figs.~\ref{fig:1dbif}(b2)--(d2) and~\ref{fig:2dbif}(b2)--(c2), such smaller extrema correspond to small oscillations around the symmetric steady state $p$, rather than full loops of $W_+^u(p)$ around either $a_-$ or $a_+$. The example in Fig.~\ref{fig:knead} demonstrates that the symbols of $S$ can be identified reliable and automatically from sufficiently long time series data of $W_+^u(p)$, which is a single trajectory that is generated by numerical integration of system~\eqref{eq1}. This allows us to compute sufficiently long kneading sequences over a fine grid in a region of interest of the $(F,\Delta)$-plane; we employ the software package Tides~\cite{tides} for this purpose.

We are interested in finding the division of parameter space into regions with the same kneading sequences. For this purpose it is very helpful to assign to each kneading sequence a single real number, called the \emph{kneading invariant}, which is then used to distinguish different kneading sequences by color. We define the kneading invariant here by 
\begin{equation}
	I =  I(S) = \sum_{k=0}^{\infty} \frac{s_k}{2^k} \in \left[0, 2 \right] \quad \text{with} \ s_k\in  \{ -1,0,+1\},
	\label{eq:knead}
\end{equation}
where the $s_k$ are given by the elements of $S$ in the obvious way. In practice, one considers the kneading invariant only up to a finite number $n$ of symbols, given by 
$$I_n = I_n(S)  = \sum_{k=0}^{n} \frac{s_k}{2^k}.$$
Since any kneading sequence generated from $W_+^u(p)$ always starts with $+$, the kneading invariant may take up to $ \sum_{i=1}^{n}2^{n-i} = 2^{n} -1$ values within the interval $\left[0, 2\right]$, which we translate via a color map to the same number of colors. 

\begin{figure}[]
	\centering
	\includegraphics[]{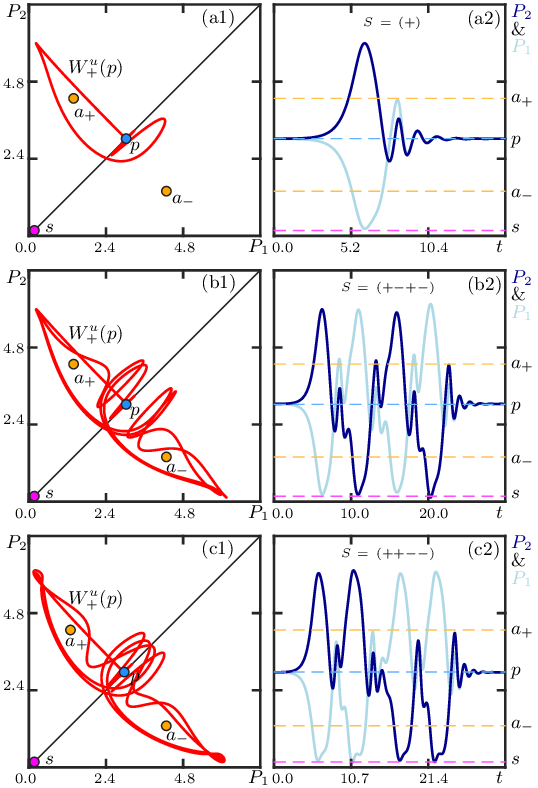}
	\caption{\label{fig:shilorb} (a1)--(c1) Three different Shilnikov homoclinic orbits formed by $W_+^u(p)$, and (a2)--(c2) corresponding temporal profiles, for $F=7.6$ and $\Delta=7.837$, $\Delta=7.897$ and $\Delta=7.624$, respectively.}
\end{figure}

As we will see in the next section, we find it useful to build up a picture of a parameter region of interest by considering the coloring by $I_n$ for different and increasing values of $n$. The reason behind this approach is that changes of $I_n$ (that is, of color) correspond to global bifurcations, here, Shilnikov bifurcations, with longer and longer excursions of $W_+^u(p)$ the larger $n$. More specifically, each such Shilnikov homoclinic orbit is characterized by a \emph{finite} kneading sequence that represents the loops that $W^u_+(p)$ makes around $a_+$ and $a_-$ before this one-dimensional manifold returns to the saddle steady state $p$. As an illustrative examples, we present in Fig.~\ref{fig:shilorb} three different homoclinic orbits formed by the branch $W_+^u(p)$; note that, due to symmetry, there also exist the corresponding homoclinic orbits formed by $W_-^u(p)$ (not shown). Panel~(a1) shows the primary Shilnikov homoclinic orbit with a single loop around $a_+$, which has the finite kneading sequence $S=(+)$ [panel~(a2)] and exists along the curve $\text{HOM}_p$ in Fig.~\ref{fig:2dbif}(a). Panels~(b1) and~(c1) of Fig.~\ref{fig:shilorb} show two different Shilnikov homoclinic orbits, where $W_+^u(p)$ completes a total of four loops before connecting back to $p$: two around $a_+$ and two around $a_-$. Panels~(b2) and~(c2) show that they feature different switching behavior, as encoded by their finite kneading sequences $S=(+-+-)$ and $S=(++--)$, respectively.

To avoid overly cumbersome notation, we do not label all the different curves of homoclinic orbits to $p$ by their finite kneading sequences in the discussion that follows. Rather, we distinguish them generally only via the overall numbers $m$ and $n$ of loops that $W_+^u(p)$ makes around $a_+$ and around $a_-$, respectively; specifically, we refer to these loci as $\text{HOM}_p^{m,n}$. For example, the curves of the homoclinic orbits in Fig.~\ref{fig:shilorb}(b) and~(c) are both labeled $\text{HOM}_p^{(2,2)}$.

\begin{figure}[]
	\centering
	\includegraphics[scale=0.95]{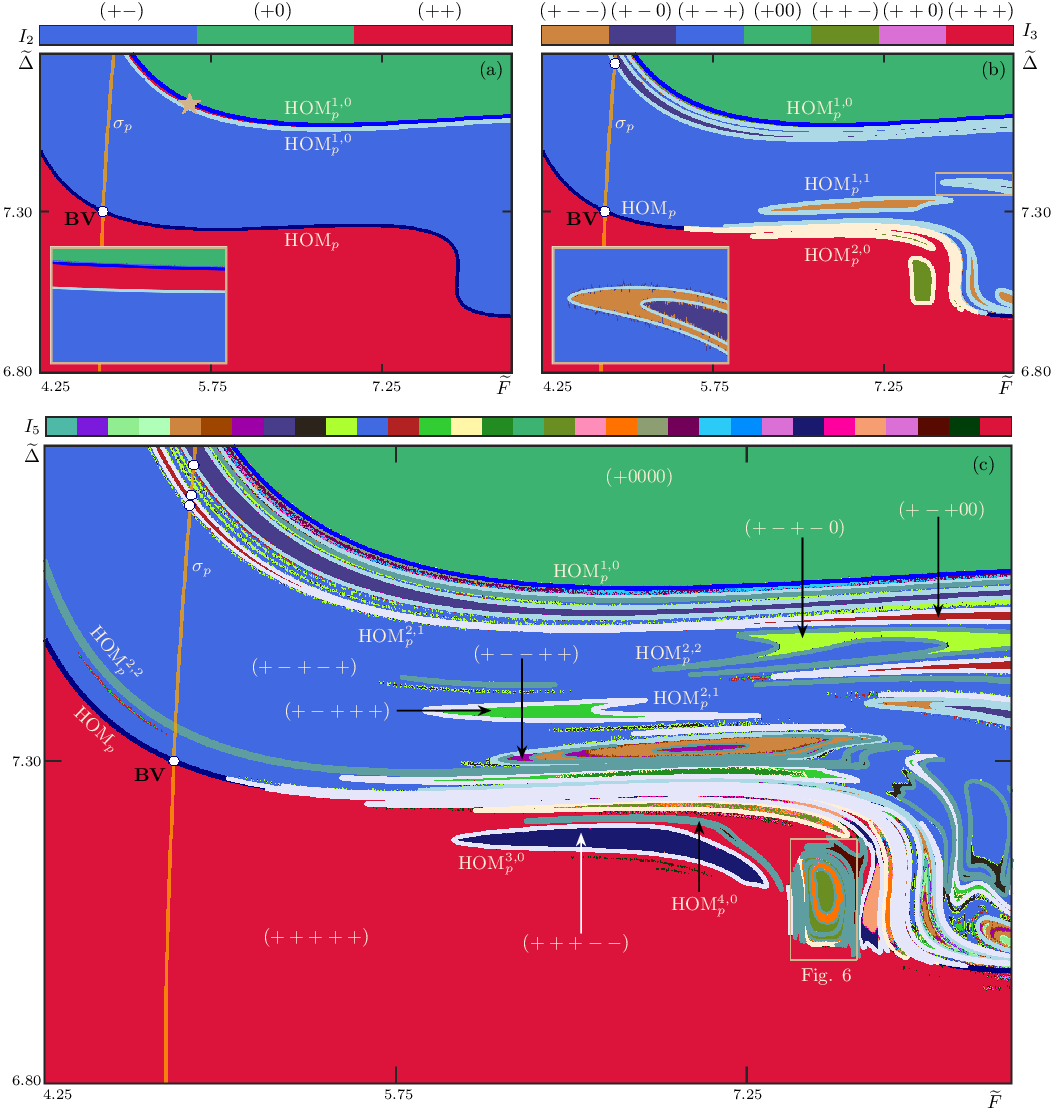}
	\caption{\label{fig:kneadmap} Kneading maps of (a) $I_2$, (b) of $I_3$, and (c) of $I_5$ in the $(\widetilde{F},\widetilde{\Delta})$-plane near the Belyakov transition point $\textbf{BV}$ of system~\eqref{eq1}, shown together with computed curves of Shilnikov bifurcations with up to four loops. Coloring distinguishes regions with different switching patterns, as indicated by the color bars.}
\end{figure}

To understand how different regions of self-switching behavior are organized, we now consider color-coded representations of the kneading invariant $I_n$ in the $(F,\Delta)$-plane for increasing $n$. More specifically, we perform parameter sweeps of the $(F,\Delta)$-plane near the point $\textbf{BV}$ over a $1000 \times 1000$ grid, and thus determine the kneading invariant $I_n$ as a function of $F$ and $\Delta$. Figure~\ref{fig:kneadmap} shows the kneading maps given by the kneading invariants $I_2$, $I_3$ and $I_5$, near the Belyakov transition point $\textbf{BV}$ in the light-brown shaded region in Fig.~\ref{fig:2dbif}(c). For its better visualisation, we mapped this region
to the $(\widetilde{F},\widetilde{\Delta})$-plane in Fig.~\ref{fig:kneadmap}, by applying a rotation over the angle $-\pi/7$ around the point $\textbf{BV}$. We proceed by discussing the kneading maps given by $I_2$, $I_3$ and $I_5$ one-by-one.

\subsection{Kneading map of $I_2$ \label{belkneadi2}}

Figure~\ref{fig:kneadmap}(a) shows the coloring by the kneading invariant $I_2$, which can and does take the three values $1/2$, $1$ and $3/2$. The $(\widetilde{F},\widetilde{\Delta})$-plane is divided into four open regions, which are colored accordingly to represent the switching patterns $(+-)$, $(+0)$, and $(++)$, as is indicated at the color bar. The large open region at the bottom with $(++)$ corresponds to the general location where we observed isolated, non-switching periodic orbits in Fig.~\ref{fig:1dbif}(b), as well as the non-switching chaotic attractor in Fig.~\ref{fig:2dbif}(b). The upper boundary of the region with $(++)$ in Fig.~\ref{fig:kneadmap}(a) is formed by the curve $\text{HOM}_p$ of the primary Shilnikov bifurcation from Fig.~\ref{fig:shilorb}(a) with the Belyakov transition point $\textbf{BV}$ on it. In the region with $(+-)$ above this curve the branch $W^u_+(p)$ performs its first two loops of oscillations around the steady states $a_+$ and $a_-$, consecutively, and then accumulates on some nearby attractor. 

As we discussed earlier, in the region with $(+0)$ at the top of Fig.~\ref{fig:kneadmap}(a), the branch $W_+^u(p)$ performs a single loop around $a_+$ and then converges to the attracting symmetric  steady state $s$. Interestingly, already on the level of $I_2$, the transition to the region with $(+-)$ is via a thin region also with $(++)$. The associated two changes of $I_2$ correspond to the two additional single-loop Shilnikov bifurcations from Fig.~\ref{fig:2dbif}(a), and we label both as $\text{HOM}^{1,0}_p$ here; see also the enlargement in the inset of Fig.~\ref{fig:kneadmap}(a) of a tiny region near the star. The fact that the Shilnikov bifurcations $\text{HOM}^{1,0}_p$ enclose an additional region with the kneading sequence $(+-)$ supports the idea that they are associated with an increase in the symmetry of associated periodic orbits and chaotic attractors.

\subsection{Kneading map of $I_3$ \label{belkneadi3}}

Considering just one extra symbol gives the kneading map of $I_3$ in  $(\widetilde{F},\widetilde{\Delta})$-plane shown in Fig.~\ref{fig:kneadmap}(b), which now features 7 colors and a substantial increase in different regions. The additional regions subdivide regions identified with $I_2$ in panel~(a); the exception is the region where $W_+^u(p)$ makes a single loop around $a_+$ and then converges to $s$, which remains unchanged and simply has a change in label from $(+0)$ to $(+00)$ in Fig.~\ref{fig:kneadmap}(b). All additional regions are delimited by new curves of non-switching and switching Shilnikov bifurcations associated with homoclinic orbits with two loops, which we also all identified and continued as curves $\text{HOM}^{2,0}_p$ and $\text{HOM}^{1,1}_p$, respectively.

Within the large region of non-switching oscillations with $(+++)$ below the curve $\text{HOM}_p$ in Fig.~\ref{fig:kneadmap}(b), one observes the presence of multiple subregions, all with the kneading sequence $(++-)$ and each bounded by a curve $\text{HOM}^{2,0}_p$. These regions and curves appear to accumulate on $\text{HOM}_p$, a considerable part of which is now obscured. When crossing into such a subregion, the output intensities $P_1$ and $P_2$ change from non-switching to switching. Hence, each Shilnikov bifurcation $\text{HOM}^{2,0}_p$ serves as a symmetry-increasing bifurcation, along which there is a merger of a pair of asymmetric chaotic attractors (or periodic orbits) with period-two-like dynamics; Fig.~\ref{fig:2dbif}(b)--(c) is actually an example of this type of transition. 

In between the curves $\text{HOM}_p$ and the top curve $\text{HOM}^{1,0}_p$ in Fig.~\ref{fig:kneadmap}(b), there is a large region with $(+-+)$ that is associated with regular switching. Within this region, we identify additional subregions with $(+--)$ and $(+-0)$, respectively, each of which is bounded by a curve of switching homoclinic bifurcations $\text{HOM}^{1,1}_p$ (with the finite kneading sequence $(+-)$). The inset of Fig.~\ref{fig:kneadmap}(b) offers a detailed view of the nested nature of two subregions in the frame on the right of the $(\widetilde{F},\widetilde{\Delta})$-plane. Notice that regions with $(+--)$ appear to accumulate on $\text{HOM}_p$ as well, now from the other side. However, there are also additional thin and stripe-like regions and associated curves $\text{HOM}^{1,1}_p$ near the boundary $\text{HOM}^{1,0}_p$ of the region with $(+00)$; these curves are not closed but effectively parallel this boundary. While this is not visible on the scale of Fig.~\ref{fig:kneadmap}(b), one finds there also several nested regions with $(++0)$, which are separated from the region with $(+++)$ by further curves $\text{HOM}^{2,0}_p$.

\subsection{Kneading map of $I_5$\label{kneadbeli5}}

When the number of kneading symbols $n$ increases, we observe a repetition of the process depicted in Figs.~\ref{fig:kneadmap}(a)--(b). Previous regions characterized by constant kneading sequences are divided further by additional curves of Shilnikov bifurcations, each with finite kneading sequences of length $n-1$. This is illustrated in Fig.~\ref{fig:kneadmap}(c), where we present the kneading map of $I_5$. With the addition of just two extra symbols that allow us to distinguish 31 regions, there is a sheer explosion of the number of subregions. We computed here also a large number of bounding curves of Shilnikov bifurcations with up to four loops, some of which are labeled.

As we already observed in panel~(b), there are many additional subregions in the lower large region of non-switching oscillation, now with $(+++++)$ in Fig.~\ref{fig:kneadmap}(c); some of these new regions are quite large, such as the one with $(+++--)$ that is bounded by a curve $\text{HOM}_p^{3,0}$. These subregions display a complicated and nested structure of accumulation onto the curve $\text{HOM}_p$ of the primary Shilnikov homoclinic orbit, as well as onto each other. Similarly, we find many additional subregions in the large region with $(+-+-+)$, which also accumulate on $\text{HOM}_p$ and on each other in an intricate way. Note that the labeled larger subregions with $(+-+++)$ and $(+-+--)$ are bounded by the curves $\text{HOM}_p^{2,1}$ and $\text{HOM}_p^{2,2}$, respectively. There are also again additional stripe-like regions and associated curves of Shilnikov bifurcations that parallel the curve $\text{HOM}_p^{1,0}$ bounding the upper region with $(+0000)$. These curves now extend further into the large region with $(+-+-+)$. Moreover, these curves cross the neutral saddle curve $\sigma_p$ and, hence, all have points of Belyakov transition. Notice that this is also the case for the curves labeled $\text{HOM}_p^{2,2}$ much closer to primary curve $\text{HOM}_p$.

The overall conclusion from Fig.~\ref{fig:kneadmap}(c) is that there is a very complicated scenario of changes to the observed switching pattern of the output light along pretty much any path that crosses the $(\widetilde{F},\widetilde{\Delta})$-plane to the right of the curve $\sigma_p$. The level of complexity is quite astounding, given that we considered only kneading sequences of length $n = 5$ here. In fact, already the kneading map of $I_5$ allows us to identify and compute a plethora of Shilnikov bifurcation curves. They provide additional and `less local' information of the effect of the Belyakov point $\textbf{BV}$ on the primary curve $\text{HOM}_p$ --- beyond the `more local' results in \cite{bitha} regarding the unfolding of a $\mathbb{Z}_2$-equivariant Belyakov transition point.

\subsection{Sensitivity of kneading sequences}

\begin{figure}[]
	\centering
	\includegraphics[]{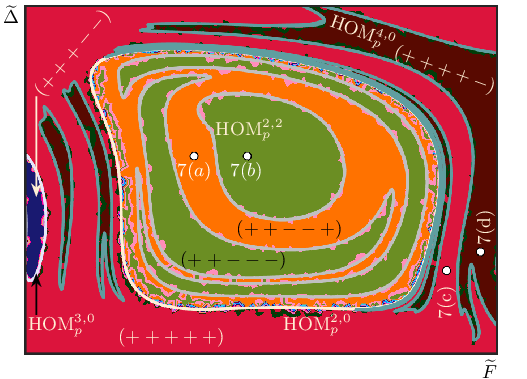}
	\caption{\label{fig:kneadmapzoom} Enlargement of the $(\widetilde{F},\widetilde{\Delta})$-plane in a region where system~\eqref{eq1} exhibits a complex arrangement of sub-regions with different kneading invariant $I_5$, as indicated by the region in the box in Fig.~\ref{fig:kneadmap}(c).}
\end{figure}

A striking feature in Fig.~\ref{fig:kneadmap}(c) are `isolas' with nested regions and closed curves of bounding Shilnikov bifurcations. An example is the isola in the shown frame, and Fig.~\ref{fig:kneadmapzoom} shows it enlarged. The island is bounded by a curve $\text{HOM}_p^{2,0}$ and consists of a region with $(++--+)$ that contains three subregions with $(++---)$, which are bounded by a curve $\text{HOM}_p^{2,2}$ of homoclinic bifurcation. This global bifurcation has the finite kneading sequence $(++--)$ and is responsible for the change of the fifth symbol, which distinguishes the two types of subregion. Figure~\ref{fig:kneadmapzoom} also features other curves $\text{HOM}_p^{3,0}$ and $\text{HOM}_p^{4,0}$, which delimit subregions with kneading sequences $(+++--)$ and $(++++-)$ respectively. Notice also the exceedingly small subregions of other colors, that is, kneading sequences, mainly near boundaries curves of larger regions. While they are at the limit of our detection algorithm, they demonstrate the acute sensitivity of the system to small parameter variations. 

\begin{figure}[]
	\centering
	\includegraphics[scale=0.95]{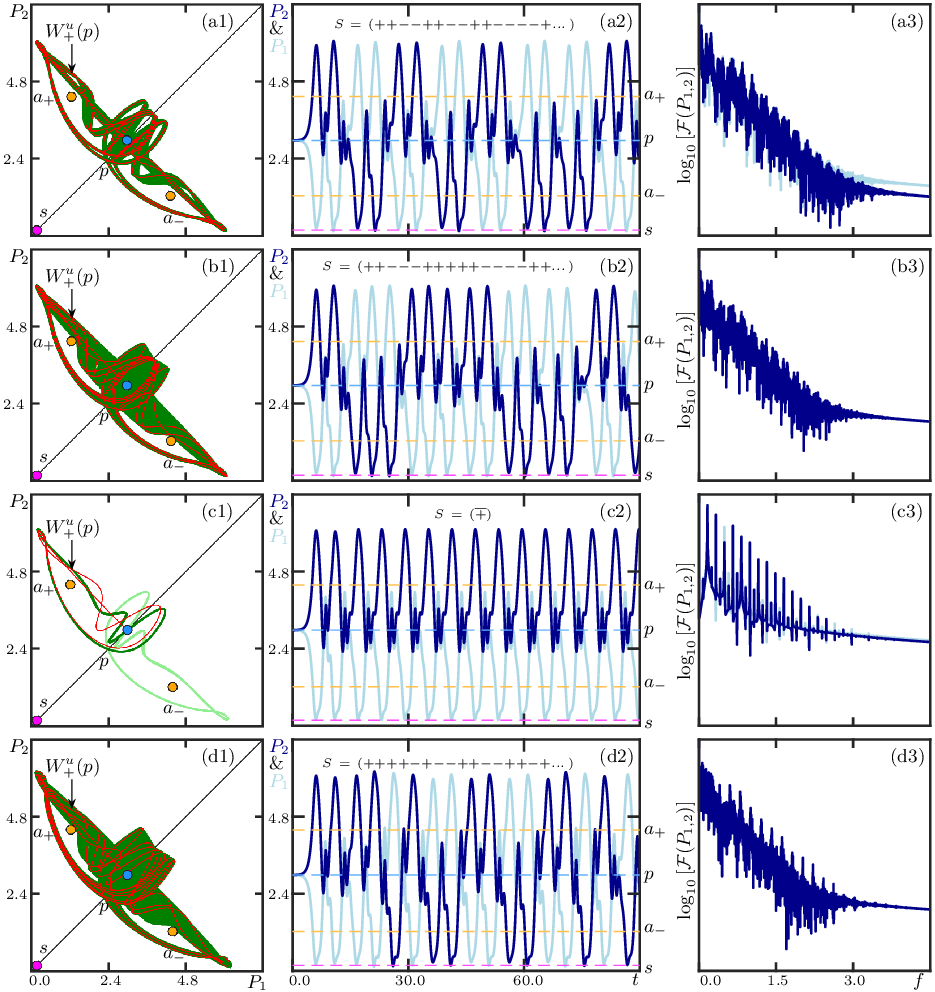}
	\caption{\label{fig:switching} (a1)--(d1) Phase portraits showing attractors with the  steady state and $W_+^u(p)$, (a2)--(d2) associated temporal traces, and (a3)--(d3) power spectra of $P_1$ and $P_2$, for the parameter points indicated in Fig.\ref{fig:kneadmapzoom}.}
\end{figure}

We now illustrate in Fig.~\ref{fig:switching} changes to the observed switching dynamics across curves $\text{HOM}_p^{2,2}$ and $\text{HOM}_p^{3,0}$, respectively, by showing phase portraits, temporal traces and power spectra at the respective pairs of points in neighboring regions that are marked in Fig.~\ref{fig:kneadmapzoom}. The transition between subregions with $(++--+)$ and $(++---)$ across $\text{HOM}_p^{2,2}$ is illustrated by rows~(a) and~(b) of Fig.~\ref{fig:switching}, and that between subregions with $(+++++)$ and $(++++-)$ across $\text{HOM}_p^{2,2}$ by rows~(c) and~(d).

At the chosen parameter point in the subregion with $(++--+)$ inside the isola bounded by a curve $\text{HOM}_p^{2,0}$ we find the chaotic attractor in Fig.~\ref{fig:switching}(a1). The corresponding time traces of $P_1$ and $P_2$ [panel~(a2)] show chaotic switching, and their power spectra are typical for a chaotic attractor. Since they are different this chaotic attractor is, in fact, asymmetric, which is not obvious from panel~(a1). In the subregion with $(++---)$ on the other side of the curve $\text{HOM}_p^{2,2}$, we also find a chaotic attractor [panel~(b1)]. Apart from the change in the fifth symbol of its kneading sequence, there does not appear to be much of a change in the nature of the time traces of $P_1$ and $P_2$ [panel~(b2)], or their power spectra [panel~(b3)]. However, these two spectra are identical, showing that this is now a symmetric chaotic attractor. Hence, the transition across the curve $\text{HOM}_p^{2,2}$ results in symmetry breaking or restoration of the chaotic attractor, depending on the direction of crossing. Notice from the complicated arrangement of the regions with $(++---)$, that one must expect several changes of the symmetry properties of the attractor along a path that crosses the isola in Fig.~\ref{fig:kneadmapzoom}.

The transition across the curve $\text{HOM}_p^{4,0}$ is of very different type. At the chosen parameter point in the region with $(+++++)$, there is the pair of attracting asymmetric periodic orbits shown in Fig.~\ref{fig:switching}(c1). The branch $W^u_+(p)$ converges to the one with dominant $P_2$ [panel~(c2)] and there is non-switching dynamics, that is, the kneading sequence is $S = (\overline{+})$.  As expected, the power spectra of $P_1$ and $P_2$ [panel~(c3)] are slightly different and show distinct harmonic frequencies. In the nearby subregion with $(++++-)$, on the other hand, we find the symmetric chaotic attractor shown in Fig.~\ref{fig:switching}(d1), with irregular switching between loops around $a_+$ and $a_-$ [panel~(d2)]. The power spectra of $P_1$ and $P_2$ [panel~(d3)] are typical and identical, and they feature remainders of the harmonic of the nearby periodic orbit.

Both transitions illustrated in Fig.~\ref{fig:switching} are quite dramatic. They concern Shilnikov bifurcations with four loops bounding subregions of the $(\widetilde{F},\widetilde{\Delta})$-plane that are distinguished by only five symbols of the kneading sequence generated by $W^u_+(p)$. The increasing complexity revealed by the successive panels of Fig.~\ref{fig:kneadmap} clearly suggests an intricate and refining picture of great complexity: these types of transitions will also occur when crossing between subregions distinguished by more symbols and bounded by associated curves of Shilnikov bifurcations with additional loops.

\section{Lyapunov map and convergence to chaotic attractors \label{lyapbelsec}}

\begin{figure}[]
	\centering
	\includegraphics[scale=0.87]{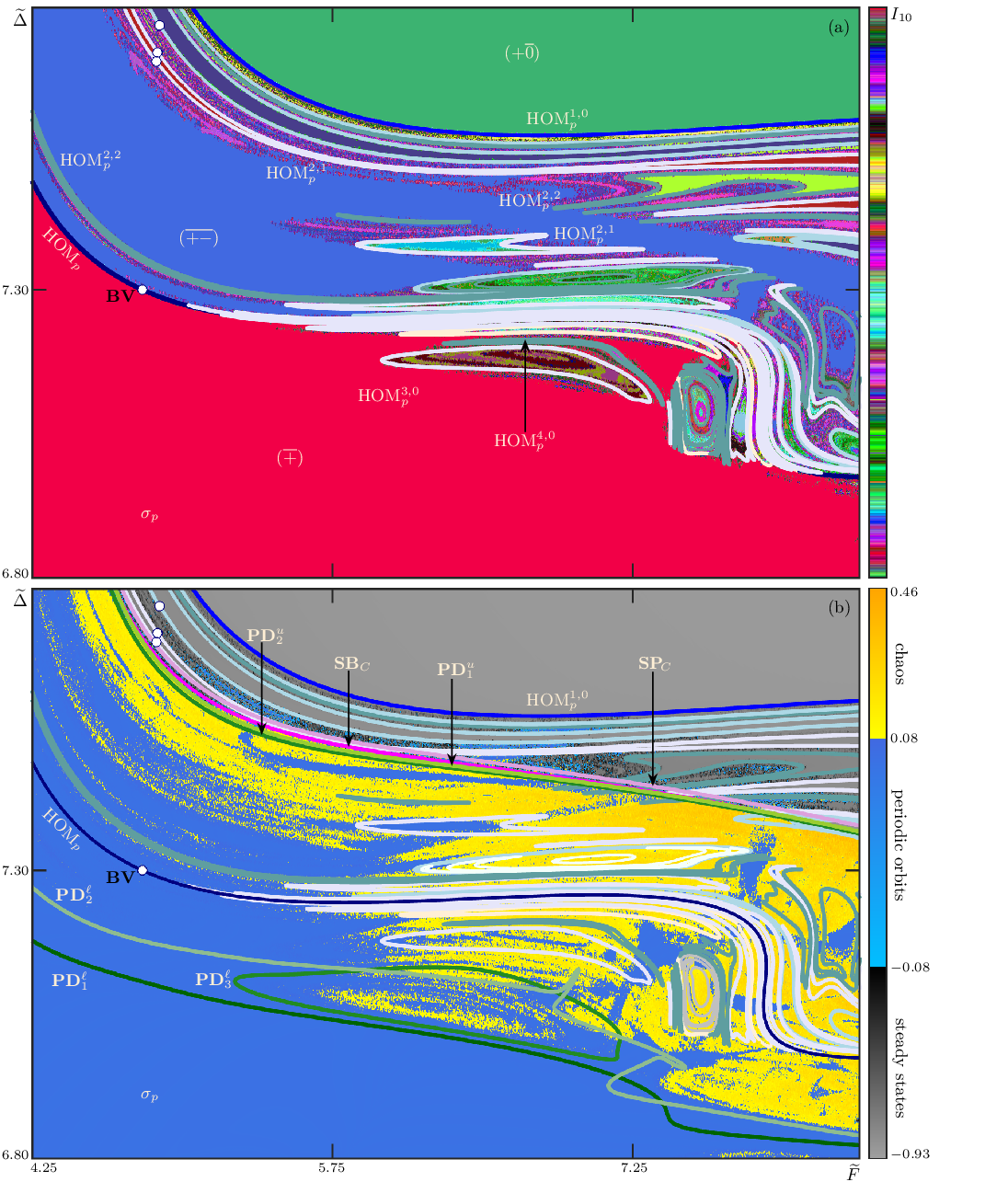}
	\caption{\label{fig:lyapunov} (a) Kneading map $I_{10}$,  and (b) maximum Lyapunov exponent of $W_+^u(p)$ in the $(\widetilde{F},\widetilde{\Delta})$-plane near the point $\textbf{BV}$ with previously shown bifurcation curves. Coloring in (b) distinguishes between  $W^u_+(p)$ converging to a steady state (gray to black), a periodic orbit (blue), or a chaotic attractor (yellow to orange); also shown are the additional curves of saddle-node $\textbf{SP}\!_C$ (light purple) and symmetry breaking  $\textbf{SB}_{C}$ (fuchsia) bifurcations of periodic orbits, and of period-doubling $\textbf{PD}_{1,2}^{u}$ and $\textbf{PD}_{1,2,3}^{\ell}$ (tones of green).}
\end{figure}

The kneading map of $I_5$ in Fig.~\ref{fig:kneadmap}(c), which features regions distinguished by only 5 symbols (of which the first one is always a $+$), already reveals a complex structure of many subregions with different kneading sequences. Subregions with kneading sequences of the form $(s_1... s_k\overline{0})$, where $W^u_+(p)$ converges to the symmetric and stable steady state $s$, are interspersed with subregions with kneading sequences, where $W^u_+(p)$ converges to either an attracting periodic orbit or to a chaotic attractor. Here, the difference in kneading sequence highlights the sensitivity of the switching behavior of system~\eqref{eq1} to parameter variations in quite large parts of the $(\widetilde{F},\widetilde{\Delta})$-plane. In this way, the kneading invariant $I_n$, even for a reasonably low number of symbols $n$, provides outstanding information of a topological nature. On the other hand, it does not actually distinguish different types of attractors. 

To address the question where which type of long-term behavior can be found, we compute the Lyapunov exponents~\cite{lyapcit} of system~\eqref{eq1} --- for consistency and definiteness, again for the branch $W^u_+(p)$. We focus here on the maximum Lyapunov exponent $\lambda_{\rm max}$, which measures the leading expansion or contraction rate and allows us to distinguish the eventual dynamics as follows. The branch $W^u_+(p)$ converges to: a steady state when $\lambda_{\rm max} < 0$; to a periodic solution when $\lambda_{\rm max} = 0$ (owing to the invariance under time shift); and to a chaotic attractor when $\lambda_{\rm max} > 0$. 

We find it very useful to complement and contrast, in the same region of parameter space, the topological information provided by the kneading map with asymptotic information on the existing attractors provided by the Lyapunov map (of the maximum exponent $\lambda_{\rm max}$). To this end, Fig.~\ref{fig:lyapunov} shows the kneading map of $I_{10}$ in panel~(a), and the Lyapunov exponent map in panel~(b), again near the Belyakov transition point $\textbf{BV}$ in the region of the $(\widetilde{F},\widetilde{\Delta})$-plane from Fig.~\ref{fig:kneadmap}(c). 

The kneading map $I_{10}$ in Fig.~\ref{fig:lyapunov}(a) is shown together with the previously presented bifurcation curves. It features many additional but increasingly smaller subregions with different switching patterns, as well as small isolas with nested subregions. However, the differences between the kneading maps of $I_{5}$ and that of $I_{10}$ are not dramatic, and this suggests that $I_{10}$ gives a sufficiently accurate representation of the limiting infinite kneading invariant $I$ from \eqref{eq:knead}. In particular, $I_{10}$ allows us to identify three main regions: that with $(+\overline{0})$, where $W^u_+(p)$ converges to the attracting  steady state $s$; that with $(\overline{+})$ of non-switching dynamics; and that with $(\overline{+-})$, where $W^u_+(p)$ has alternating excursions around $a_+$ and $a_-$. 

The corresponding Lyapunov map is shown in Fig.~\ref{fig:lyapunov}(b). Here, the maximum Lyapunov exponent was computed over a $1000 \times 1000$ grid by the established method of integrating the linearization of system~\eqref{eq1} along the trajectory $W^u_+(p)$, here over $1000$ units of time, while using a continuous Gram–Schmidt orthonormalization process to control the components of the Lyapunov vectors; see Ref.~\cite{lyapcit} for details. To account for the tolerances of this computation based on taking long-term averages, we make the following practical distinction represented by the color bar: we identify a steady state when $\lambda_{\rm max} < -0.08$; a periodic orbit when $\lvert \lambda_{\rm max} \lvert < 0.08$; and a chaotic attractor when $0.08 < \lambda_{\rm max}$; these two boundaries were found by numerical experimentation as suitable for resolving the dynamics with sufficient accuracy. Also shown here are again all previous bifurcation curves, as well as additional curves of saddle-node bifurcation, symmetry breaking and period-doubling of periodic orbits that are relevant for the onset of chaotic dynamics.

Comparison between panels~(a) and~(b) of Fig.~\ref{fig:lyapunov} shows that the Lyapunov map and the kneading map concern related yet different information. Together, they are able to provide a comprehensive picture of a complex dynamical landscape when complemented with relevant curves of bifurcations. This general approach has been used to map out simple versus chaotic dynamics in other dynamical systems; in particular, Refs.~\cite{andrus2021,barrio,barrio2} are in this spirit. There is some broadbrush agreement between where in the $(\widetilde{F},\widetilde{\Delta})$-plane one finds strong sensitivity of the switching dynamics with many subregions of $I_{10}$ in panel~(a), and where chaotic dynamics is identified in panel~(b). However, there are indeed also considerable differences. The upper curve $\text{HOM}^{1,0}_p$  clearly lies in the region where $\lambda_{\rm max} < -0.08$; hence, it is definitely not the boundary of the region of the $(\widetilde{F},\widetilde{\Delta})$-plane where chaotic dynamics can be found. Rather, we observe a sudden transition to chaotic dynamics along the shown curve $\textbf{SB}_C$ of symmetry-breaking to a symmetric periodic orbit. More precisely, the pair of asymmetric periodic orbits that emerges from $\textbf{SB}_C$ undergoes a saddle-node bifurcation $\textbf{SP}\!_C$, followed almost immediately by a cascade of period-doubling bifurcations leading to chaos. Figure~\ref{fig:lyapunov}(b) shows the curve $\textbf{SP}\!_C$ just above the curve $\textbf{SB}_C$, with the first two period-doubling bifurcations $\textbf{PD}^u_1$ and $\textbf{PD}^u_2$ that are also very close to the curve $\textbf{SB}_C$. Below these upper boundary curves, the branch $W^u_+(p)$ converges to a chaotic attractor in large areas of the $(\widetilde{F},\widetilde{\Delta})$-plane but, as expected from theory, these are `interrupted' by windows where $W^u_+(p)$ converges to an attracting periodic orbit. Notice that many closed curves of Shilnikov bifurcations `align' with certain areas of chaotic dynamics, while other closed curves clearly also contain larger areas of periodic dynamics. This is particularly prominent above and below the primary curve $\text{HOM}_p$ well to the right of the point $\mathbf{BV}$, where chaotic dynamics appears to be prominent. 

The lower boundary in the $(\widetilde{F},\widetilde{\Delta})$-plane of where chaotic dynamics can be found is much less `crisp' compared to the upper boundary. There is not a `uniform' cascade of period-doublings over the shown $\widetilde{F}$-range. Rather, we find a more complicated arrangement of curves of period-doubling bifurcations, represented in Fig.~\ref{fig:lyapunov}(b) by the three curves $\textbf{PD}^{\ell}_1$,  $\textbf{PD}^{\ell}_2$ and $\textbf{PD}^{\ell}_3$. They roughly `align' with certain subareas where a chaotic attractor can be found, and the curve $\textbf{PD}^{\ell}_1$ provides a practical lower boundary. 

It is clear from Fig.~\ref{fig:lyapunov} that system~\eqref{eq1} exhibits further global bifurcations in the region between the curves $\textbf{SB}_C$ and $\textbf{PD}^{\ell}_1$, which account for the observed subtleties in the sign of the maximum Lyapunov exponent $\lambda_{\rm max}$. Moreover, these curves and the area with chaotic attractors in between them clearly extend to higher values of $\widetilde{F}$, that is, higher input power $F$. An investigation of how the bifurcation diagram is organized there will be reported elsewhere.

\section{Conclusion and outlook \label{con4}}

We presented a detailed analysis of various types of self-switching oscillations in a single-mode ring resonator with interacting light modeled by system~\eqref{eq1}, where we focused on the parameter region near a point of a Belyakov transition on a curve of primary homoclinic bifurcations. In order to characterize different types of self-switching dynamics, we computed symbolic information concerning the behaviour of the positive branch $W^u_+(p)$ of the unstable manifold of the symmetric saddle steady state $p$. We presented this information in the form of the finite kneading maps showing the kneading invariant $I_n$ for an increasing number $n$ of symbols. This staged approach allowed us to identify systematically a complex sequence of non-switching and switching Shilnikov bifurcations, each responsible for creating periodic orbits or chaotic attractors with different switching patterns. Specifically, we observed that the transition of periodic orbits or chaotic attractors through each of these global bifurcations leads to the emergence of new oscillating solutions characterized by different symmetry properties. The latter is directly related to the finite switching pattern of the respective Shilnikov homoclinic orbit whose locus bounds corresponding subregions of $I_n$ with `neighboring' symbols.

The kneading map of $I_{10}$, the most refined one we computed, was then compared with with the Lyapunov map showing the maximum Lyapunov exponent. In this way, we were able to complement and contrast the switching properties of system~\eqref{eq1} with information on where which types of attractors can be found. While the areas with chaotic attractors largely coincide with regions of strong parameter sensitivity of the switching behavior, there is no one-to-one correspondence between the two. In fact, in a considerable part of the parameter plane, there is strong switching sensitivity without chaotic dynamics; rather the sensitivity is due to subregions of eventual periodicity, which are interspersed with subregions with convergence to a steady-state.

Overall, we found an intriguing interplay between different types of switching behaviors. Our investigation shed further light on the nature of the $\mathbb{Z}_2$-equivariant Belyakov transition~\cite{bitha}, and how bifurcation curves emerging from it generate complicated periodic and chaotic switching dynamics. While fine details of the dynamics may not be accessible experimentally, we expect that the broader aspects of out finding are within experimental reach~\cite{selfswi, bruno2020}, provided the system is symmetrically balanced. This concerns, in particular, larger regions of regular switching, and the distinction between those being periodic or chaotic in nature. The high sensitivity of the switching behavior during transitions from one pattern to another may also be observable along suitably chosen paths in parameter space.


\begin{acknowledgments}
R. D. Dikand{\'e} Bitha acknowledges the Korea Institute for Advanced Study for providing the opportunity for a research visit, during which part of this work was done. A. Giraldo was supported by the KIAS Individual Grant No. CG086102 at the Korea Institute for Advanced Study.

\end{acknowledgments}

\appendix


\bibliography{BGBK_bel_kneading}

\end{document}